\documentclass[prl,twocolumn,showpacs,amsmath,amssymb]{revtex4}
\usepackage{graphicx}
\usepackage{dcolumn}
\usepackage{bm}

\begin{document}

\title{\boldmath
Precison Measurements of the Mass, the Widths of  $\psi(3770)$ Resonance 
and the Cross Section $\sigma[e^+e^-\rightarrow \psi(3770)]$
at $E_{\rm cm}=3.7724$ GeV}
 \author{
\begin{small}
M.~Ablikim$^{1}$,      J.~Z.~Bai$^{1}$,            Y.~Ban$^{11}$,
J.~G.~Bian$^{1}$,      X.~Cai$^{1}$,               H.~F.~Chen$^{15}$,
H.~S.~Chen$^{1}$,      H.~X.~Chen$^{1}$,           J.~C.~Chen$^{1}$,
Jin~Chen$^{1}$
% ...
\end{small}
\\(BES Collaboration)\\
}
\affiliation{ 
\begin{minipage}{145mm}
$^{1}$ Institute of High Energy Physics, Beijing 100049, People's Republic
of China\\
% ...
$^{11}$ Peking University, Beijing 100871, People's Republic of China\\
% ...
$^{15}$ University of Science and Technology of China, Hefei 230026,
People's Republic of China\\
% ...
\end{minipage}
}
\begin{abstract}
By analyzing the $R$ values measured at 68 energy points in the energy
region between 3.650 and 3.872 GeV reported in our previous paper,
we have precisely measured
the mass, the total width, the leptonic width and the 
leptonic decay branching fraction of the $\psi(3770)$
to be ${M}_{\psi(3770)}=3772.4 \pm 0.4 \pm 0.3$ MeV,
$\Gamma_{\psi(3770)}^{\rm tot} = 28.6 \pm 1.2 \pm 0.2$ MeV,
$\Gamma_{\psi(3770)}^{ee} = 279 \pm 11 \pm 13$ eV
and $B[\psi(3770)\rightarrow e^+e^-]=(0.98\pm 0.04\pm 0.04)\times 10^{-5}$,
respectively,
which result in 
the observed cross section
$\sigma^{\rm obs}[e^+e^-\rightarrow \psi(3770)]=7.25\pm 0.27 \pm 0.34$ nb
at $\sqrt{s}=3772.4$ MeV.
We have also measured
$R_{\rm uds}=2.121\pm 0.023 \pm 0.084$ for the continuum light hadron
production
in the region from 3.650 to 3.872 GeV.
\end{abstract}
\pacs{13.85.Lg, 12.38.Qk, 14.40.Gx, 14.40.Lb}

\maketitle

Precise measurements of the mass ${M}_{\psi(3770)}$ , the total width
$\Gamma_{\psi(3770)}^{\rm tot}$ and the leptonic
width $\Gamma_{\psi(3770)}^{ee}$ of the $\psi(3770)$ resonance,
and the cross section $\sigma[e^+e^-\rightarrow \psi(3770)]$
for the $\psi(3770)$ production 
at $\sqrt{s}=M_{\psi(3770)}$ 
in the $e^+e^-$ annihilation
are important in understanding the nature of the $\psi(3770)$. 
The $\psi(3770)$ is thought to be a maxiture of the $D$-wave and $S$-wave
of the angular momentum eignstates of $c\bar c$ system. 
The detailed mixing scheme affacts the
$\psi(3770)$ production and decays. 
So precise measurements of 
these quantities
would give us some useful information about the nature of
the $\psi(3770)$
and offer some insights into the internal wave functions of the charmonium,
which are beneficial to understand the dynamics of the $1^{--}$ resonance
production in the $e^+e^-$ annihilation.
Potential or quarkonium
models based on QCD can calculate the
masses~\cite{eichten_prl34_y1975_p369,eichten_prd21_y1980_p203,
heikkila_prd29_y1983_p110}
of the $c \bar c$ bound states,
the total widths of the $\psi(3770)$ and other resonances
~\cite{eichten_prd21_y1980_p203,heikkila_prd29_y1983_p110},
and
the width for $\psi(3770)\rightarrow D\bar D$
~\cite{eichten_prd69_y2004_p094019,heikkila_prd29_y1983_p110}.
The Lattice QCD (LQCD) can calculate the 
mass spectra of
the $Q\bar Q$ system (heavy quark and anti-quark system,
such as the $c\bar c$ and $b\bar b$ system)~\cite{prd52_y1995_p6519_davies}. 
If one can more precisely measure the masses and the widths 
of the $Q\bar Q$ system, the results of these measurements 
can be used to test the calculation of the quantities 
by the models and by the LQCD theory.
Moreover, these measurements can in turn be used
to extract two fundamental parameters in QCD, the $c$-quark mass and
the strong coupling constant $\alpha_s(s)$ at
this mass scale~\cite{prd52_y1995_p6519_davies}. 

     In recent days, 
new results of the measured quantities releated to the $\psi(3770)$
production and decays were
reported~\cite{bes_dd_nondd_psipp,cleo_dd_nondd_psipp}. 
These improve our knowledge on charmonium production and decays,
especially improve our understanding of 
the nature of the $\psi(3770)$ resonance.
In our previous work~\cite{rplxx_y2006_pyyyy_bes} we have reported
measurements of the $R(s)$ values measured
at 68 energy points in the region between 3.650 and 3.872 GeV,
where the quantity $R(s)$ is defined as the ratio  
$\sigma(e^+e^- \rightarrow {\rm hadrons})/(\sigma(e^+e^- \rightarrow \mu^+\mu^-)$,
with $\sigma(e^+e^- \rightarrow\mu^+\mu^-)=3\pi {\alpha}^2(0)/4s$,
here $s$ is the c.m. (center-of-mass) energy squared and
$\alpha(0)$ is the fine structure constant at the lowest energy limit.
In this Letter we report the results obtained by further analyzing
these $R(s)$ values.
From this analysis we obtain 
${M}_{\psi(3770)}$, $\Gamma_{\psi(3770)}^{\rm tot}$,
$\Gamma_{\psi(3770)}^{ee}$
and
the leptonic branching fraction for $\psi(3770)\rightarrow e^+e^-$
with improved precision 
compared to those of the PDG~\cite{pdg} world average, 
and we obtain the cross section 
$\sigma[e^+e^-\rightarrow \psi(3770)\rightarrow {\rm hadrons}]$ at 
$\sqrt{s}=3772.4$ MeV
with a precision better than
any of those measured in cross section scan experiments previously.  

In Ref.~\cite{rplxx_y2006_pyyyy_bes} we reported measurements of the quantities
$R_{\rm uds}$, $R_{\rm had}(s)$ and $R_{{\rm uds(c)}+\psi(3770)}$
which are the
$R(s)$ values for the continuum light hadron production 
around the $D\bar D$ threshold, 
$R(s)$ values including the contributions
from the continuum hadrons and all
$1^{--}$ resonances at all energies,
and the $R(s)$ values accounting for the contributions from
both the continuum hadron production and the decays for
$\psi(3770)\rightarrow {\rm hadrons}$, respectively.
All of these are corrected for the intial state radiative and
vacuum polarization corrections. 
To extract the mass and widths of
the $\psi(3770)$,
we here, in the Letter, analyze the
quantity $R_{{\rm uds(c)}+\psi(3770)}$~\cite{rplxx_y2006_pyyyy_bes}.
Table~\ref{67pnts_r} summarizes the $R_{{\rm uds(c)}+\psi(3770)}$ values
reported in Ref.~\cite{rplxx_y2006_pyyyy_bes}.

\begin{table*}[t]
\caption{Summary of the $R_{\rm uds(c)+\psi(3770)}(s)$ values measured at 68 energy points.
}
\label{67pnts_r}
\begin{center}
\begin{tabular}{c c c c c c c c} \hline \hline
 $\sqrt{s}$ & $R_{\rm uds(c)+\psi(3770)}$ &$\sqrt{s}$ & 
$R_{\rm uds(c)+\psi(3770)}$ & $\sqrt{s}$ & $R_{\rm uds(c)+\psi(3770)}$ &
$\sqrt{s}$  & $R_{\rm uds(c)+\psi(3770)}$   \\ 
 (GeV)  &   &  (GeV)  &      &   (GeV)  &    &  (GeV) &    \\ \hline
3.6500 & $2.157\pm 0.035\pm 0.086$ &
       3.7584 & $3.025\pm 0.108\pm 0.148$   &
              3.7726 & $3.777\pm 0.145\pm 0.185$ &
                     3.7826 & $3.326\pm 0.115\pm 0.163$  \\
3.6600 & $2.131\pm 0.105\pm 0.085$ &
       3.7596 & $3.076\pm 0.102\pm 0.151$   &
              3.7730 & $3.563\pm 0.120\pm 0.175$ &
                     3.7838 & $3.154\pm 0.114\pm 0.155$  \\
3.6920 & $2.034\pm 0.092\pm 0.081$   &
       3.7608 & $3.138\pm 0.089\pm 0.154$   &
              3.7742 & $3.373\pm 0.113\pm 0.165$ &
                     3.7850 & $2.879\pm 0.107\pm 0.141$  \\
3.7000 & $2.079\pm 0.079\pm 0.083$   &
       3.7620 & $2.992\pm 0.110\pm 0.147$   &
              3.7754 & $3.641\pm 0.125\pm 0.178$ &
                     3.7862 & $2.902\pm 0.105\pm 0.142$  \\
3.7080 & $2.197\pm 0.083\pm 0.088$   &
       3.7622 & $3.207\pm 0.114\pm 0.157$   &
              3.7766 & $3.498\pm 0.119\pm 0.171$ &
                     3.7874 & $2.957\pm 0.111\pm 0.145$  \\
3.7160 & $2.177\pm 0.086\pm 0.087$   &
       3.7634 & $3.345\pm 0.122\pm 0.164$   &
              3.7778 & $3.570\pm 0.121\pm 0.175$ &
                     3.7886 & $2.571\pm 0.097\pm 0.126$  \\
3.7240 & $2.125\pm 0.086\pm 0.085$   &
       3.7646 & $3.585\pm 0.126\pm 0.176$   &
              3.7790 & $3.360\pm 0.117\pm 0.165$ &
                     3.7898 & $2.576\pm 0.099\pm 0.126$  \\
3.7320 & $2.156\pm 0.086\pm 0.086$   &
       3.7658 & $3.381\pm 0.119\pm 0.166$   &
              3.7798 & $3.477\pm 0.136\pm 0.170$ &
                     3.7900 & $2.849\pm 0.106\pm 0.140$  \\
3.7400 & $2.190\pm 0.099\pm 0.088$   &
       3.7670 & $3.760\pm 0.130\pm 0.184$   &
              3.7802 & $3.427\pm 0.125\pm 0.168$ &
                     3.7950 & $2.751\pm 0.101\pm 0.135$  \\
3.7480 & $2.371\pm 0.106\pm 0.116$   &
       3.7682 & $3.451\pm 0.124\pm 0.169$   &
              3.7804 & $3.382\pm 0.137\pm 0.166$ &
                     3.8000 & $2.212\pm 0.091\pm 0.108$  \\
3.7500 & $2.517\pm 0.085\pm 0.123$   &
       3.7694 & $3.611\pm 0.125\pm 0.177$   &
              3.7808 & $3.336\pm 0.129\pm 0.163$ &
                     3.8100 & $2.171\pm 0.092\pm 0.087$  \\
3.7512 & $2.637\pm 0.090\pm 0.129$   &
       3.7706 & $3.580\pm 0.123\pm 0.175$   &
              3.7810 & $3.464\pm 0.138\pm 0.170$ &
                     3.8200 & $2.367\pm 0.109\pm 0.095$  \\
3.7524 & $2.615\pm 0.095\pm 0.128$   &
       3.7714 & $3.538\pm 0.139\pm 0.173$   &
              3.7812 & $3.396\pm 0.130\pm 0.166$ &
                     3.8300 & $2.354\pm 0.101\pm 0.094$  \\
3.7536 & $2.652\pm 0.093\pm 0.130$   &
       3.7716 & $3.634\pm 0.146\pm 0.178$   &
              3.7814 & $3.514\pm 0.124\pm 0.172$ &
                     3.8400 & $2.296\pm 0.104\pm 0.092$  \\
3.7548 & $2.733\pm 0.093\pm 0.134$   &
       3.7718 & $3.939\pm 0.133\pm 0.193$   &
              3.7816 & $2.944\pm 0.137\pm 0.144$ &
                     3.8500 & $2.372\pm 0.115\pm 0.095$  \\
3.7560 & $2.585\pm 0.090\pm 0.127$   &
       3.7720 & $3.636\pm 0.134\pm 0.178$   &
              3.7818 & $3.140\pm 0.125\pm 0.154$ &
                     3.8600 & $2.371\pm 0.105\pm 0.095$  \\
3.7572 & $2.942\pm 0.107\pm 0.144$   &
       3.7722 & $3.652\pm 0.143\pm 0.179$   &
              3.7822 & $3.253\pm 0.124\pm 0.159$ &
                     3.8720 & $2.308\pm 0.117\pm 0.092$  \\
\hline \hline
\end{tabular}
\end{center}
\end{table*}

\begin{figure}
\includegraphics[width=7.0cm,height=6.5cm]
{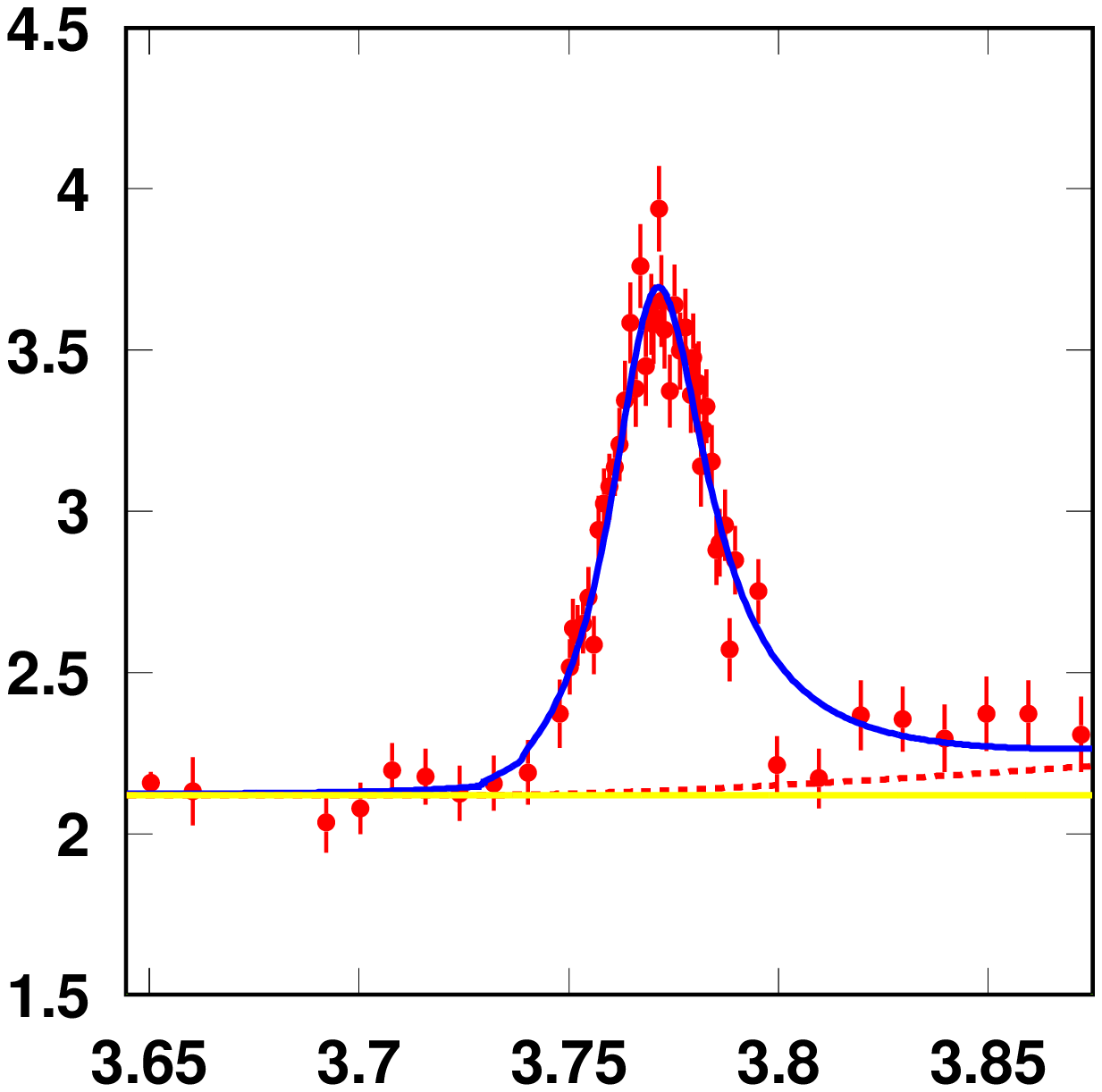}
\put(-115.0,-10.0){\bf{$E_{\rm cm}$ [GeV]}}
\put(-210,56){\rotatebox{90}{\bf $R_{\rm uds(c) +\psi(3770)}(s)$}}   
\caption{The $R_{\rm uds(c) +\psi(3770)}(s)$ versus the c.m. energy (see
text).
}
\label{r_67pnts_fit}
\end{figure}

The determination of 
${M}_{\psi(3770)}$, $\Gamma_{\psi(3770)}^{\rm tot}$
and $\Gamma_{\psi(3770)}^{ee}$
is accomplished by simultaneously fitting
the measured $R_{{\rm uds(c)}+\psi(3770)}$ values 
listed in table~\ref{67pnts_r} to the function
that describes the combined $\psi(3770)$ resonance shape and
non-resonant hadronic background.
Assuming that there are no other new structures and effects,
we use a pure p-wave zero order Breit-Wigner function
with energy-dependent total widths
to describe the $\psi(3770)$ production
and its decay to inclusive hadrons.
The $\psi(3770)$ resonance shape is taken as
{\begin{small}
\begin{equation}
R_{\psi(3770)}(s) =\frac{3s}{4\pi \alpha^2(0)}
    \frac{12 \pi \Gamma^{0~ee}_{\psi(3770)}
            \Gamma_{\psi(3770)}(s)}
{{(s-M_{\psi(3770)}^2)^2 +
[M_{\psi(3770)}\Gamma_{\psi(3770)}(s)]^2}},
\end{equation}
\end{small} } 
\noindent
\hspace{-3.8mm} with 
$\Gamma^{0~ee}_{\psi(3770)}=|1-\Pi(s(1-x))|^2\Gamma^{ee}_{\psi(3770)}$,
where $\Gamma^{0~ee}_{\psi(3770)}$ and $\Gamma^{ee}_{\psi(3770)}$
are the bare leptonic width excluding the vacuum polarization effects
and experimental leptonic width including the vacuum polarization effects,
respectively;
$1/{|1-\Pi(s(1-x))|^2}$ is the vacuum polarization correction
function~\cite{berends}
including the contributions from all
$1^{--}$ resonances,
the QED continuum hadron spectrum
as well as the contributions from the lepton pairs ($e^+e^-$,
$\mu^+\mu^-$ and $\tau^+\tau^-$)~\cite{zhangdh_gen};
$x$ is a parameter related to the total
energy of the emitted photons;
the total width 
$\Gamma_{\psi(3770)}(s)$
is chosen to be energy dependent defined as
{\begin{small}
\begin{equation}
 \Gamma_{\psi(3770)}(s)=\Gamma_{D^0\bar D^0}(s)+
          \Gamma_{D^+D^-}(s)+\Gamma_{{\rm non}-D\bar D}(s),
\end{equation}
\end{small} } 
\hspace{-0.32cm} in which~\cite{prl97_y2006_p121801}
\begin{eqnarray}
\Gamma_{D^0\bar D^0}(s) & = &
\Gamma^{\rm tot}_{\psi(3770)}\theta_{00}
           \frac{(p^{}_{D^0})^3} {(p^0_{D^0})^3}
            \frac{1+(rp_{D^0}^0)^2} {1+(rp_{D^0})^2}B_{00},
\end{eqnarray}
\begin{eqnarray}
\Gamma_{D^+D^-}(s) & = &
\Gamma^{\rm tot}_{\psi(3770)}\theta_{+-}
           \frac{(p^{}_{D^+})^3} {(p^0_{D^+})^3}
            \frac{1+(rp_{D^+}^0)^2} {1+(rp_{D^+})^2}B_{+-},
\end{eqnarray}
and
\begin{equation}
\Gamma_{{\rm non-}D\bar D}(s) = 
\Gamma^{\rm tot}_{\psi(3770)}\left[ 1 - B_{00} - B_{+-}\right],
\end{equation}
where $p^0_D$ and $p_D$ are
the momenta of the $D$ mesons
produced at the peak of the $\psi(3770)$ and at
the c.m. energy $\sqrt{s}$, respectively;
$\Gamma^{\rm tot}_{\psi(3770)}$
is the total width of
the $\psi(3770)$ at its peak,   
$B_{00}$ and $B_{+-}$
are the branching fractions for
$\psi(3770)\rightarrow D^0\bar D^0$ and
$\psi(3770)\rightarrow D^+D^-$, respectively,
$r$ is the interaction radius of the $c\bar c$,
$\theta_{00}$ and $\theta_{+-}$ are the step   
functions to account for the thresholds of the $D^0\bar D^0$ and
$D^+D^-$ production, respectively. 
In the fit we take 
$\Gamma^{\rm tot}_{\psi(3770)}$
and
$r$ as free parameters.
Since we do not select the $D\bar D$ events from these data samples, we can
not leave $B_{00}$ and $B_{+-}$ free in the fit. Instead we
set the ratio 
$B_{00}/B_{+-}=1.3\pm0.1$~\cite{prl97_y2006_p121801,npb727_y2005_p395} 
and the 
branching fraction for $\psi(3770)\rightarrow D\bar D$ to be
$B[\psi(3770) \rightarrow D\bar D]=B_{00}+B_{+-}=
(84 \pm 7)\%$~\cite{prl97_y2006_p121801,r_and_bf} 
in the fit.

The non-resonant background shape is taken as
{\small
\begin{eqnarray}
R_{\rm uds(c)}(s) = R_{\rm uds}(s) + R_{(c)}(s),
\end{eqnarray}
}
\noindent
with
{\small
\begin{eqnarray}
R_{(c)}(s) = f_{D\bar D}\left[ (\frac{p_{D^0}}{E_{D^0}})^3\theta_{00}
               + (\frac{p_{D^+}}{E_{D^+}})^3\theta_{+-} \right],
\end{eqnarray}
}
\noindent
\hspace{-3.5mm}
where 
$R_{\rm uds}(s)$ is the $R$ value for the continuum light hadron production;
$R_{(c)}(s)$ is the $R$ value due to the continuum 
$D \bar D$ ($D^0 \bar D^0$ and $D^+D^-$) production;
$E_{D^0}$ and $E_{D^+}$ are the energies of the $D^0$ and the $D^+$ mesons
produced at 
$\sqrt{s}$, respectively; 
$f_{D\bar D}$ is a parameter to be fitted. 

The expected $R_{{\rm uds(c)}+\psi(3770)}(s)$ value is written as
\begin{eqnarray}
\nonumber  R^{\rm expect}_{{\rm uds(c)}+\psi(3770)}(s) & =
                  \int_{0}^{\infty}ds'' G(s,s'')
            R_{\psi(3770)}(s) & \\
   & \hspace{-1.6cm} +~~R_{\rm uds(c)}(s),
\end{eqnarray}
\noindent
\hspace{-0.11cm} in which $G(s,s'')$ is the Gaussian function to describe the c.m. energy
distribution of the BEPC machine~\cite{prl97_y2006_p121801}.
In the fit we leave $R_{\rm uds}(s)$
free, assuming that it is independent of the energy.

Figure~\ref{r_67pnts_fit} shows 
the measured $R_{{\rm uds(c)}+\psi(3770)}(s)$ values 
with the fit, where the point with errors represents 
the $R_{{\rm uds(c)}+\psi(3770)}(s)$ value,
the curve (blue line) gives the fit to the data,
the straight (yellow) line shows the quantity $R_{\rm uds}$ obtained from the
fit and
the dashed (red) line shows the variation of
the quantity $R_{\rm uds(c)}(s)$ with the c.m. energy 
as given in Eqs. (6-7).
The $\chi^2/ndf$ for
this fit is 
$94/61=1.5$.
The larger $\chi^2$s are mainly from the four energy points, which are
at $\sqrt{s}=3.7886, ~3.7898, ~3.80$ and $3.81$ GeV. 
The sum over the $\chi^2$s
from these four energy points is 31. 
If we exclude 
these four points in the fit,
we would obtain $\chi^2/ndf=65/57=1.1$.
The "dip" of the $R_{{\rm uds(c)}+\psi(3770)}(s)$ values around 3.80 GeV,
which obviously turn aside from the expected ones,
may be due to some unknown 
effects
~\footnote{
The analysis author would like to suggest 
the BES-III (BES-III at BEPC-II) 
and/or CLEO (CLEO-c at CERS) Collaborations
to check whether this "dip" is due to physical reason
with larger statistical cross section scan data to be collected
at more sampling points around $\sqrt{s}=3.80$ GeV
in the future.
}.
Table~\ref{tbl_psipp_prmt} summarizes the results from the fit, 
where the first error is the
statistical and the second systematic.
The systematic errors on $\Gamma^{\rm tot}_{\psi(3770)}$
and $\Gamma^{ee}_{\psi(3770)}$ arise 
mainly from the common systematic uncertainty
on the measured quantity $R_{{\rm uds(c)}+\psi(3770)}(s)$, while
the systematic error on ${M}_{\psi(3770)}$ arises 
from the uncertainty in the energy calibration~\cite{prl97_y2006_p121801}
of the BEPC collider.
As a comparison, the same quantities measured by the BES Collaboration
obtained by analyzing different cross section scan 
data sets~\cite{prl97_y2006_p121801}
are also listed in the table.
The fit also gives
$$R_{\rm uds}=2.121\pm 0.023 \pm 0.084$$
in the energy region from 3.650 to 3.872 GeV, where the errors are
statistical and systematic, respectively. The systematic uncertainty arises
mainly from the uncertainty ($\sim 3.9\%$) in the measurements of 
$R_{{\rm uds(c)}+\psi(3770)}(s)$. 
The measured $R_{\rm uds}$ value from this fit is consistent with the
$R_{\rm uds}=2.141\pm 0.025 \pm 0.085$ obtained by weighting the first 8
$R_{{\rm uds(c)}+\psi(3770)}(s)$ values below the $D\bar D$
threshold~\cite{rplxx_y2006_pyyyy_bes}.
In estimating the systematic uncertainties,
the effects of the uncertainty in the
measurement of the branching fraction for $\psi(3770)\rightarrow D\bar D$
on the measured quantities are also taken into account. 
The measured widths of the $\psi(3770)$ yield its leptonic branching
fraction shown in table~\ref{tbl_psipp_lptnc_bf}.

\begin{figure}
\includegraphics[width=8.0cm,height=8.0cm] 
{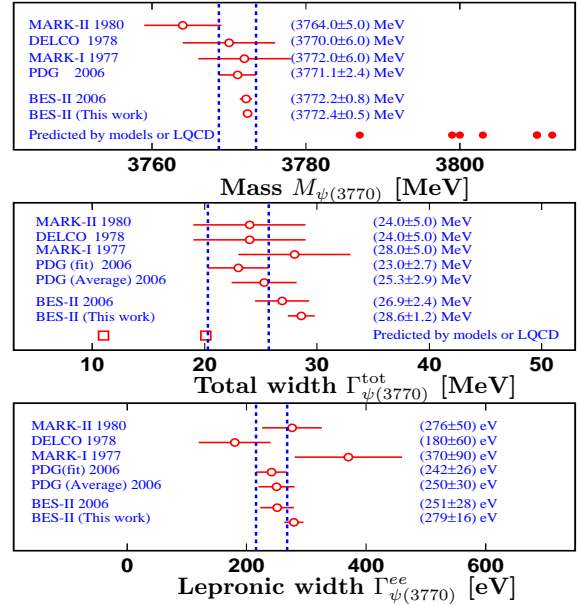}
\put(-142.0,148.0){\bf Mass {$M_{\psi(3770)}$ [MeV]}}
\put(-152.0,73.0){\bf Total width {$\Gamma^{\rm tot}_{\psi(3770)}$ [MeV]}}
\put(-158.0,-3.0){\bf Lepronic width {$\Gamma^{ee}_{\psi(3770)}$ [eV]}}
\caption{Comparison of the measured quantities obtained
by analyzing the cross section scan data  
from different experiments along with those predicted by the models and 
by the LQCD theory; 
the dots are the predicted mass of the $1^3D_1$ state of the $c \bar c$
system and the squares the expected partial width for
$\psi(3770)\rightarrow D \bar D$ (see text).
}
\label{psi3770_prmt_cmprsn}
\end{figure}

\begin{table}
\centering   
\caption{    
The measured mass ${M}_{\psi(3770)}$, 
total width $\Gamma^{\rm tot}_{\psi(3770)}$
and leptonic width $\Gamma^{ee}_{\psi(3770)}$ of the $\psi(3770)$
along with those measured previously by the BES
Collaboration~\cite{prl97_y2006_p121801}.}
\label{tbl_psipp_prmt}
\begin{tabular}{cccc} \hline \hline
 ${M}_{\psi(3770)}$~(MeV) & $\Gamma^{\rm tot}_{\psi(3770)}$~(MeV) &
$\Gamma^{ee}_{\psi(3770)}$~(eV) & Note \\ \hline
$3772.4 \pm 0.4 \pm 0.3$  & $28.6\pm 1.2 \pm 0.2$ &
$279 \pm 11 \pm 13$ & This work \\
$3772.2 \pm 0.7 \pm 0.3$  & $26.9\pm 2.4 \pm 0.3$ &
$251 \pm 26 \pm 11$ & \cite{prl97_y2006_p121801}\\
\hline \hline
\end{tabular}
\end{table}  
\begin{table}
\centering
\caption{The leptonic branching fraction of the $\psi(3770)$.}
\label{tbl_psipp_lptnc_bf}
\begin{tabular}{cc} \hline \hline
 Experiment & $B[\psi(3770) \rightarrow e^+e^-]$~ $\times 10^{-5}$ \\ \hline
This work   &   $0.98 \pm 0.04 \pm 0.04$                           \\
BES~\cite{prl97_y2006_p121801} &   $0.93\pm 0.06 \pm 0.03$         \\
PDG~\cite{pdg}                 &   $1.05\pm 0.14$     \\
\hline \hline
\end{tabular}
\end{table}
Figure~\ref{psi3770_prmt_cmprsn} illustrates the comparisons of the
measured quantities (circle with error bar) 
with those measured by
the MARK-I, DELCO and MARK-II
Collaborations~\cite{mark-i,delco,mark-ii} 
from analyzing the cross section scan data. 
As a comparison we also plot the
mass (dots) of the $1^3D_1$ state of the $c \bar c$ system 
and the partial width
(squares~\cite{heikkila_prd29_y1983_p110,eichten_prd69_y2004_p094019}) 
for $\psi(3770)\rightarrow D \bar D$
predicted by the 
models~\cite{eichten_prd21_y1980_p203,heikkila_prd29_y1983_p110,
modes_expct_mass_widths} in recent years
and calculated by the LQCD theory~\cite{prd52_y1995_p6519_davies} 
in the figure.

With the mass, total width and leptonic width of the $\psi(3770)$, 
we extract the $\psi(3770)$ production
cross section $\sigma^{\rm prd}_{\psi(3770)}$ at $\sqrt{s}=M_{\psi(3770)}$,
which excludes the initial state radiative corrections,
and extract its corresponding observed cross section 
$\sigma^{\rm obs}_{\psi(3770)}$,
which includes the initial state radiative  
and vacuum polarization effects.
Table~\ref{tbl_xsct_psi3770} 
summarizes the 
measured cross sections from this measurement and those measured by
the MARK-II Collaboration and by the BES Collaboration 
from analysis of different cross section scan data~\cite{prl97_y2006_p121801}.
\begin{table}
\centering   
\caption{The cross sections for the $\psi(3770)$ production measured 
at $\sqrt{s}=M_{\psi(3770)}$,
obtained by analyzing the cross section scan data over the $\psi(3770)$ or
over both the $\psi(3770)$ and $\psi(3686)$.
}
\label{tbl_xsct_psi3770}
\begin{tabular}{cccc} \hline \hline
Experiment & $\sigma^{\rm prd}[e^+e^-\rightarrow \psi(3770)]$~ &
~$\sigma^{\rm obs}[e^+e^-\rightarrow \psi(3770)]$ \\ 
           & [nb]         &         [nb]             \\ \hline
This work                           & $10.06\pm 0.37 \pm 0.043$
                                    & $7.25\pm 0.27 \pm 0.34$  \\
BES~\cite{prl97_y2006_p121801}  & $9.63\pm 0.66\pm 0.35$ 
                                    & $6.94\pm0.48 \pm0.28$        \\
MARK-II~\cite{mark-ii} &              & $9.3 \pm 1.4$~\cite{mark-ii} \\
\hline \hline
\end{tabular}
\end{table}  

In summary, 
we measured the mass of the $\psi(3770)$ with a precision of 
more than a factor of 4 better than the one of the PDG~\cite{pdg} average.
The precise measurement of the mass
combining with the measured masses
of other $c\bar c$ states could be used to gauge the effect of systematic
improvement on the calculation of the $c \bar c$ spectrum and
to extract the $\alpha_s(s)$ at this mass
scale~\cite{prd52_y1995_p6519_davies}.
Our measured 
$\Gamma^{\rm tot}_{\psi(3770)}=28.6 \pm 1.2 \pm 0.2$ MeV 
is significantly larger than the PDG average
$\Gamma^{\rm tot}_{\psi(3770)}=23.0 \pm 2.7$ MeV~\cite{pdg},
and 
with a precision more than a factor of 2 better
than that of
the PDG average. It is also obviously larger 
than the recently resonable prediction
by the coupled channel
models~\cite{heikkila_prd29_y1983_p110,eichten_prd69_y2004_p094019}
by more than $7\sigma$, where $\sigma$ is the error of the measured value
of the total width of $\psi(3770)$.
Our measured leptonic width of the $\psi(3770)$ 
is consistent within error with the PDG average~\cite{pdg},
but with a better accuracy than the PDG average.
The measured leptonic branching fraction of the $\psi(3770)$
from this work is consistent within error with PDG average and with that
measured by the BES Collaboration previously~\cite{prl97_y2006_p121801}, 
but with a precision of
more than a factor of 2 better than
the PDG average.
We measured the $\psi(3770)$ production cross section
to be 
$\sigma^{\rm obs}_{\psi(3770)} = 7.25\pm 0.27 \pm 0.34$ nb
at $\sqrt{s}=M_{\psi(3770)}$.
These improved measurements of the quantities
would be helpful for us to understand 
the nature of the $\psi(3770)$ resonance.  
From the analysis we also extracted 
$R_{\rm uds}=2.121\pm 0.023 \pm 0.084$
in the energy range from 3.650 to 3.872 GeV.
This information would be beneficial for us in the understanding of 
the continuum hadron production in or nearby the resonance region(s).
The measured $R_{\rm uds}$ can directly be used to extract the $\alpha_s(s)$
at this energy scale.
\vspace{0.5mm}
 
   The BES collaboration thanks the staff of BEPC for their hard efforts.
This work is supported in part by the National Natural Science Foundation
of China under contracts
Nos. 19991480,10225524,10225525, the Chinese Academy
of Sciences under contract No. KJ 95T-03, the 100 Talents Program of CAS
under Contract Nos. U-11, U-24, U-25, and the Knowledge Innovation Project
of CAS under Contract Nos. U-602, U-34(IHEP); by the
National Natural Science
Foundation of China under Contract No.10175060(USTC),and
No.10225522(Tsinghua University).

\end{document}